\documentclass[aps,pre,onecolumn,superscriptaddress,10pt,floatfix]{revtex4}
\usepackage[dvips]{graphics}
\usepackage{graphicx}
\usepackage{graphics}
\usepackage{amsfonts}
\usepackage{amssymb}
\usepackage{amsmath}
\usepackage{subfigure}
\usepackage{amsmath}
\usepackage{epstopdf}
\usepackage{tikz}
\usepackage{circuitikz}

\DeclareGraphicsExtensions{.pdf,.eps,.png,.jpg,.mps}
\DeclareMathOperator{\sech}{sech}

\DeclareMathOperator{\arcsinh}{arcsinh}

\begin{document}
\title{From Solitons to Rogue Waves in Nonlinear Left-Handed Metamaterials}

%\author{Yannan Shen}
%\affiliation{Department of Mathematics and Statistics, California State University, Northridge, CA, 91330, USA }

\author{Yannan Shen}
\affiliation{Department of Mathematics and Statistics, California State University, Northridge, CA, 91330, USA }

\author{P.G.\ Kevrekidis}
\affiliation{Department of Mathematics and Statistics, University of Massachusetts, Amherst MA 01003-4515, USA}
\author{G.P. Veldes} 
\affiliation{Department of Physics, National and Kapodistrian University of Athens, Panepistimiopolis, Zografos, Athens 15784, Greece}
\affiliation{Department of Electronics Engineering, Technological Educational Institute of Central Greece, Lamia 35100, Greece}

\author{D.J.\ Frantzeskakis}
\affiliation{Department of Physics, National and Kapodistrian University of Athens, Panepistimiopolis, Zografos, Athens 15784, Greece}

\author{D. DiMarzio}
\affiliation{NG Next, Northrop Grumman Aerospace Systems, One Space Park, Redondo Beach, CA 90278 USA}

\author{X. Lan  }
\affiliation{NG Next, Northrop Grumman Aerospace Systems, One Space Park, Redondo Beach, CA 90278 USA}

\author{V. Radisic}
\affiliation{NG Next, Northrop Grumman Aerospace Systems, One Space Park, Redondo Beach, CA 90278 USA}

\begin{abstract}
In the present work, we explore soliton and rogue-like wave solutions 
in the transmission line analogue of a nonlinear left-handed metamaterial. 
The nonlinearity is expressed through
a voltage-dependent and symmetric capacitance 
motivated by the
recently developed ferroelectric barium strontium titanate (BST) 
thin film capacitor designs.  We develop both the corresponding nonlinear dynamical 
lattice, as well as its reduction via a multiple scales expansion
to a nonlinear Schr{\"o}dinger (NLS) model for the envelope of a given carrier
wave. The reduced model can feature either a focusing or a defocusing
nonlinearity depending on the frequency (wavenumber) of the carrier. 
%wave. 
We then consider the robustness of different types of solitary
waves of the reduced model within the original nonlinear left-handed medium.
%Initially, 
We find that both bright and dark solitons 
%solitary waves (solitons) 
persist in a suitable parametric regime, where the reduction to the NLS is valid. 
%For related media that are also described by suitable NLS equations, rogue-like Peregrine solitons have been predicted and observed.  
%For the nonlinear left handed transmission line under consideration, 
Additionally, for suitable initial conditions, we observe a rogue
%, Peregrine-like
wave type of behavior, 
that %also 
differs significantly from the classic Peregrine rogue wave evolution, including most notably 
the breakup of a single Peregrine-like pattern into solutions with multiple wave peaks.
%The significance of this is discussed in relation to the specific nonlinearity introduced via the BST capacitance.
Finally, we touch upon the behavior of 
generalized members of the family of the Peregrine solitons, namely Akhmediev
breathers and Kuznetsov-Ma solitons, and explore how these evolve in the left-handed transmission line.
%context of the NLTL.
\end{abstract}

\maketitle

\section{Introduction}

%Metamaterials are artificial, effectively homogeneous structures, featuring negative refractive index at specific frequency bands where the effective permittivity ε and permeability μ are simultaneously negative [1, 2]. In contrast to ordinary materials which are composed by atoms, metamaterials are constructed by structural components called meta-atoms [3].

Over the past few years, the study of metamaterials, i.e., 
artificially engineered structures exhibiting electromagnetic (EM) properties 
not commonly observed in nature, has seen an explosion of interest 
%as is evidenced by numerous reviews/books~
\cite{review1,review2,review3,review4}.
An especially intriguing aspect of these metamaterials is their so-called 
left-handed (LH) nature, which features
simultaneously negative effective permittivity $\epsilon$ and permeability
$\mu$, i.e., the relevant signs of these quantities are opposite to
those of conventional right-handed (RH) media.
The resulting difference between these two scenarios
is that in the LH (RH) regime,
the energy and the wave fronts of the EM waves
propagate in %the 
opposite (same) directions, giving rise to backward- (forward-)
propagating waves.
Consequently, these left handed metamaterials (LHM) can exhibit negative
refraction at microwave \cite{exp1,exp2} or %even 
optical frequencies \cite{shalaev}.

Apart from a classical EM 
%electromagnetic 
approach involving the study of an effective medium, which can naturally be used
to study such metamaterial media~\cite{kong}, transmission line (TL) theory
constitutes a convenient framework to analyze their evolutionary
dynamics. A TL-based analysis relies on the connection of the EM 
%electromagnetic 
properties of the medium ($\epsilon$ and $\mu$)
with the electric elements of the TL unit cell, namely, the serial and
shunt impedance~\cite{review2}. Equivalent TL models have been used to
describe periodic lattices of prototypical magnetic and electric
metamaterial structures in the form of split ring resonators (SRRs)
and complementary split ring resonators (CSRRs)~\cite{ref6,ref7}.
In this context, each of the SRRs/CSRRs can be analyzed in the form
of a corresponding LC circuit, while the whole metamaterial is an array
of such circuits, with the coupling between the elements
being modeled by a mutual inductance / capacitance. The serial and shunt
impedance are directly related to the actual properties of these structures.

In addition to the more standard case of linear LHMs, the
study of nonlinear LHMs has been receiving increased
attention \cite{yurirmp}. Here, the EM properties -- such as $\epsilon$ and $\mu$
(or, equivalently, the serial and shunt impedance at the TL level) -- 
depend on the intensity of the EM field (equivalently
at the TL level, e.g., on the voltage). Practical proposals
for the experimental realization of such features
involve embedding an array of wires and SRRs into
a nonlinear dielectric~\cite{zharov,agranovich}, or
the insertion of diodes (varactors) into resonant
conductive elements, such as the SRRs~\cite{lapine,soukoulis,ysk}.
The interplay of strong dispersion exhibited by left handed transmission lines with the nonlinear voltage dependence of the group velocity results in unusual
dynamical behavior. 
In this theme, the extensive theoretical studies have
led to numerous experimental realizations of features
such as pulse propagation \cite{koz2},
envelope soliton formation \cite{koz3} and the emergence
of bright \cite{ogas} or dark \cite{wang} solitons;
see also Ref.~\cite{ourlars}, and the more recent work \cite{ofyexp} for soliton generation 
in active metamaterials. A relevant -- but earlier -- review of experimental studies
can be found in Ref.~\cite{revtl}.

In our present considerations, we study 
%examine 
a nonlinear left handed transmission line. In part, we are motivated specifically by the recent development 
of strongly nonlinear and voltage symmetric barium strontium titanate (BST) thin film 
capacitors~\cite{Meyers}. We thus consider a
nonlinear LHM through a TL approach, which exhibits the symmetric capacitance-voltage dependence.
 Our aim is to investigate 
%try to understand 
the properties of the nonlinear waveforms that arise and are robustly
sustained by this LHM. To gain theoretical insight into this, we utilize a
%multiple scales 
multiscale expansion method that reduces 
%transforms 
the model, in a self-consistent
fashion (up to cubic order in a suitable amplitude parameter), 
to a nonlinear Schr{\"o}dinger (NLS) equation~\cite{sulem,ablowitz,siambook}.
We identify regimes, depending on 
%as a function of 
the frequency of the carrier wave, where the NLS equation 
%in the considered dynamics the nature of the latter equation, i.e., whether it 
is focusing or defocusing. The prototypical soliton
solutions of this model, namely the bright and the dark
soliton, are found to be robustly preserved by the transmission line
dynamics. However, a more ambitious goal of the present study is
to examine whether {\it rogue wave} (RW) patterns, such as the Peregrine
soliton (PS) of the focusing NLS equation, can emerge 
%are preserved 
in the LH transmission line. The only work that we are aware of connecting these
two themes (LHMs and RWs) is that of~\cite{kofane}, which focuses on 
%only presents 
a rather qualitative comparison for very short propagation
distances. Here, we actually {\it engineer} initial data that,
at the NLS level, would lead to a PS profile. As a result
of the dynamics, we observe both similarities with and differences
from what we expect at the NLS level. We discuss these
at some length and the impact that the intrinsic
features of the LHM system have on the potential emergence
and form of the Peregrine-like structure. We do not restrict
our considerations to solely this rogue wave pattern; rather,
we extend them to additional members of the relevant family of
solutions, including the spatially periodic Akhmediev breathers (ABs) and the
temporally periodic Kuznetsov-Ma solitons (KMs).

%The subsequent 
Our presentation is structured as follows. In section II we present the model, 
discuss the nature of the nonlinearity and explain the reduction to the NLS setting. 
In section III, we present prototypical numerical results not only for the bright 
and dark solitons but also, more importantly, for Peregrine-like solitons/rogue waves 
and related (periodic in space or time) patterns.
Finally, in section IV, we summarize our findings and present our conclusions.

\section{The model and its analytical consideration}
%Theoretical Reduction}
\begin{figure}[h]
\includegraphics[width=8cm]{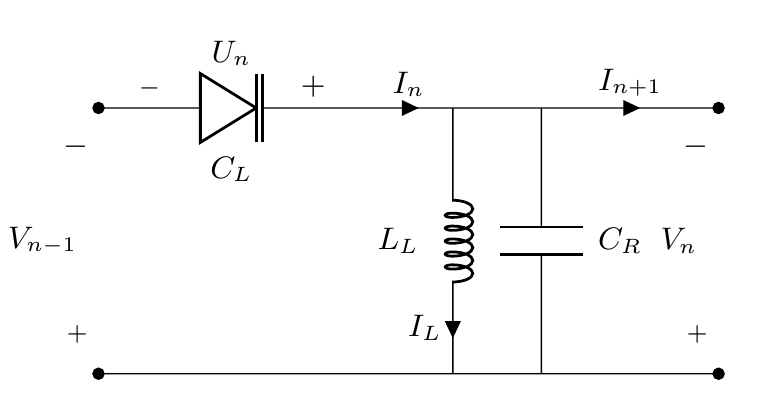}
\caption{A sketch of the left-handed transmission line emulating
the left handed metamaterial.}
\label{expsys}
\end{figure}

%\begin{figure}[h]
%\begin{center}
%\begin{circuitikz}[american voltages]
%\draw[scale=0.9]
%  (-1,0) to [short, *-] (4,0)
%  to [C, l_=$C_R$] (4,3) 
%  to [short,] (3,3) 
%  (4,0) to [short] (6,0)
%  (6,0) to [short, *-] (6,0)
%  to [open, v^<=$V_{n}$] (6,3)
%  to [short, *- ] (6,3)
%  to [short, i_<=$I_{n+1}$ ] (4,3)
%  (-1,0) to [open, v^<=$V_{n-1}$] (-1,3) 
%  to [short, *- ] (-1,3) 
%  %to [L, l_=$L_R$] (1,3)
%  to [VCo, l_=$C_L$, v^>=$U_n$] (2,3) 
%  to [short,i^=$I_n$] (3,3) 
%  to [ L, l_=$L_L$, i_>=$I_L$,] (3,0); 
%  \end{circuitikz}
%  \caption{A sketch of the left-handed transmission line emulating
%the left handed metamaterial.}
% \end{center}
% \label{expsys}
%\end{figure}

%In the earlier work of
Following Ref.~\cite{ourlars}, we consider the transmission line framework of Fig.~1
%~\ref{expsys}
in order to model 
%from first principles (i.e., Kirchhoff's
%voltage and current laws) an electrical lattice emulating 
a left handed TL metamaterial. Assuming that the diode can be 
emulated by a nonlinear, voltage-dependent capacitance (see below), we can 
%medium.
employ Kirchhoff's voltage and current laws for the unit-cell circuit 
of Fig.~1,
%~\ref{expsys},
and derive (see details in Ref.~\cite{ourlars}) the following 
differential-difference equation for the unknown voltage $V_n$ in 
the parallel branch at the $n$-th site of the lattice:
%%%%%%%%%%%%%%%%%%%%%%%%%%%%
%The resulting model in that case will be the starting point
%of our present considerations in the form of the dynamical equations
%(for the notes $n$ of the lattice):
\begin{eqnarray}
%&&
L_L \frac{d^2}{dt^2}[C_L(U_n)U_n - C_L(U_{n+1})U_{n+1}]
%\nonumber \\
%&-&
-L_LC_R \frac{d^2 V_n}{dt^2}-V_n=0.
\label{int2}
\end{eqnarray}
Here, $U_n$ is the voltage across the nonlinear capacitance $C_L$ that emulates 
the in-line BST capacitor 
(notice that $U_n=V_{n-1}- V_n$), $C_R$ is the linear shunt capacitance, while 
$L_L$ represents the inductive elements, connected
to the ground; notice that subscripts $L$ and $R$ denote the LH and RH elements 
of the unit cell circuit.
%, while $C_L$ the
%capacitive ones in the transmission line. The latter correspond
%to an in-line BST capacitor.
%Notice that $U_n=V_{n-1}- V_n$.
A key feature considered herein 
is the {\it symmetric} (contrary to what was the case
in Ref.~\cite{ourlars}) dependence of the capacitive element
on the voltage $V$ of the form:
\begin{eqnarray}
C_L(U) = \frac{C_0}{2 \cosh(\frac{2}{3} \arcsinh(\frac{2 V}{V_0})) - 1} \approx C_0-\frac{16C_0}{9V_0^2} U^2 +O(U^4),
\label{taylor2}
\end{eqnarray}
%Again this was 
motivated by results in strongly nonlinear voltage symmetric capacitance 
for molecular beam epitaxy grown BST thin films~\cite{Meyers}. For the voltages, we will assume for simplicity a Taylor
expansion and an accurate to O$(U^4)$ functional form
provided in the second expression of Eq.~(\ref{taylor2}). Here $C_0$ is the zero bias capacitance and $V_0$ is the ``2:1'' voltage, i.e., $C(V_0) = C_0/2$.  We assume the the 
fringing capacitance of~\cite{Meyers} is vanishing herein.

As a result of this expansion, Eq.~(\ref{int2}) becomes:
\begin{eqnarray}
%&&
L_L C_0 \frac{d^2}{dt^2}(V_{n-1}-2V_n+V_{n+1})-L_L C_R \frac{d^2 V_n}{dt^2}-V_n
%\nonumber \\
%&+&
+L_L \frac{d^2}{dt^2}\left\{ a [(V_{n-1}-V_n)^3 - (V_n-V_{n+1})^3] \right\}=0.
\label{mod1}
\end{eqnarray}
where $a = (16/9)(C_0/V_0^2)$. 
%$a = \frac{16C_0}{9U_0^2}$. 
Next, measuring time in units of $1/\omega_0 =\sqrt{L_L C_0}$ and voltage in units of $3V_0/4$, 
we express Eq.~(\ref{mod1}) in the following dimensionless form: 
\begin{eqnarray}
%&&
\frac{d^2}{dt^2}(V_{n-1}-2V_n+V_{n+1})-g \frac{d^2 V_n}{dt^2}-V_n
%\nonumber \\
%%&-&\alpha \frac{d^2}{dt^2}\left\{[(V_{n-1}-V_n)^2 - (V_n-V_{n+1})^2] \right\}
%\nonumber \\
%&+&\
%+\gamma 
+\frac{d^2}{dt^2}\left\{ [(V_{n-1}-V_n)^3 - (V_n-V_{n+1})^3] \right\}=0,
\label{modp}
\end{eqnarray}
where $g=C_R/C_0$. 
%and $\gamma  = a V_0^2/C_0 = (16/9)(V_0^2/U_0^2)$.
%$g = \frac{C_R}{C_0}$, and $\gamma  = \frac{a V_0^2}{C_0} = \frac{16V_0^2}{9U_0^2}$.

To obtain an analytical handle on the nonlinear waveforms
that the model of Eq.~(\ref{modp}) may possess, we will employ
a quasi-continuum approximation \cite{Rem}. In particular, we consider
waveforms 
characterized by a discrete carrier and a slowly-varying continuum pulse-like envelope, 
and thus seek solutions of Eq.~(\ref{modp}) of the form: 
\begin{equation}
V_n =\sum_{\ell=1} \epsilon^{\ell} V_{\ell}(X,T)e^{i\ell (\omega t - k n)} +{\rm c.c.},
%\qquad \theta_n = \omega t - k n,
\label{eq:ansatz}
\end{equation}
where $V_{\ell}$ are unknown envelope functions, depending on the slow variables:
\begin{equation}
X=\epsilon (n-v_g t), \qquad T = \epsilon^2 t,
\label{slow}
\end{equation}
with $v_g$ being the group velocity, as can be found self-consistently 
from the linear dispersion relation (see below). 
Finally, $\omega$ and $k$ denote the carrier's frequency and wavenumber, respectively, 
and $\epsilon$ is a formal small parameter. 

%In practical terms, 
Here we should notice that the above ansatz implies that we are assuming a carrier 
wave of effective linear propagation, as will be more transparent
in what follows. This carrier wave is modulated by a slow envelope
that encompasses the nonlinear dynamics of the model. This slow envelope
is expected to be governed, as we will see in what follows, by the
NLS model. This expansion is a small-amplitude one (i.e., weakly nonlinear), 
%expansion 
as the relevant control parameter $\epsilon$ characterizes
the solution amplitude. At the same time, it is a long-wavelength
expansion characterizing wide regions of the lattice of size of
$1/\epsilon$ and long time scales of the size of $1/\epsilon^2$.

We now %examine 
present the resulting equations from the multiscale expansion
order by order.
%\begin{eqnarray}
%O(\epsilon ^{1})&:& [-1 + (2 + g- 2 \cos k) \omega^2 ] \big(e^{i(\omega t - k n) }V_1 + 
%  e^{-i(\omega t - k n)}V_1^*\big ) = 0,    \qquad
%%i.e. \quad\frac{1}{ \omega^2 }  =  2 + g- 2 \cos k;
%\nonumber\\
%O(\epsilon ^{2})&:& (-\frac{v_g}{\omega}+\omega^2 \sin k ) \big[2 i e^{i(\omega t - k n) }\partial_X V_1 -2 i e^{-i(\omega t - k n)}\partial_XV_1^*\big] + \nonumber\\
%&& e^{2i(\omega t - k n) } [16 i \alpha \omega^2  (\cos k -1) \sin k  {V_1}^2  + (3+8\omega^2 (\cos k -\cos(2k)))V_2] + \nonumber\\
%&& e^{-2i(\omega t - k n)}[-16 i \alpha \omega^2 (\cos k -1)\sin k {V_1^*}^2+(3+8\omega^2 (\cos k -\cos(2k)))V_2^*) ]= 0,    \nonumber\\
%%&& i.e.\quad {v_g} = - \omega^3 \sin k ;\qquad V_2 = \frac{-16 i \alpha \omega^2  (\cos k -1) \sin k  {V_1}^2}{3+8\omega^2 (\cos k -\cos(2k))}\\
%O(\epsilon ^{3})&:& e^{i(\omega t - k n)}\{[- 4i  +2i e^{-ik}+2i e^{ik}-2i g ]\omega\partial_T V_1 +\nonumber\\ &&[-2v_g^2+e^{-ik}v_g^2+e^{ik}v_g^2-g v_g^2-2i e^{-ik}\omega v_g+2ie^{ik} \omega v_g -\frac{1}{2} e^{-ik} \omega^2 -\frac{1}{2} e^{ik} \omega^2 ]\partial_X^2 V_1 \}  \nonumber\\
% &&+ e^{2i(\omega t - k n)}\{ \alpha\omega (\omega + 2 \omega \cos(k) + 2 v \sin(k)) V_1\partial_X V_1[-8e^{-ik}+16-8e^{ik} +4i \omega (v (2 + g - 2 \cos(2 k)) + 2 \omega\sin(2 k)) \partial_X V_2
% \}\nonumber\\
% &&+e^{3i(\omega t - k n)}\{ -144\gamma\omega^2 (1 + 2 \cos(k)) \sin^4(k/2) V_1^3 +\alpha\omega^2 (\sin(k) + \sin(2 k)) V_1 V_2 [36i e^{-ik}-72i +36 i e^{ik}] \nonumber\\ 
% &&- (1 - 9 (2 + g) \omega^2 + 18 \omega^2 \cos(3 k)) V_3 \} +c.c.
%\end{eqnarray}
%i.e. 
\begin{eqnarray}
O(\epsilon ^{1})&:& \frac{1}{ \omega^2 }  =  2 + g- 2 \cos k; \label{disp} \\
O(\epsilon ^{2})&:&  {v_g} = - \omega^3 \sin k ;\qquad V_2 = 0;\\
O(\epsilon ^{3})&:& i \partial_T V_1+P \partial_{X}^2 V_1 + Q |V_{1}|^2 V_{1}=0,
\quad P = \frac{\omega^3}{2} (\cos k-3 \omega^2\sin^2 k ), 
\quad Q = -24
%\gamma 
\omega^3\sin^4(k/2);
\label{exp}
\\
&& V_3 = \frac{144
%\gamma 
\omega^2 (1+2\cos k ) \sin^4(k/2) V_1^3}{1+g-2\cos(3k)} \nonumber;  
\end{eqnarray}

\begin{figure}[tbp]
\centering
\includegraphics[scale=0.4]{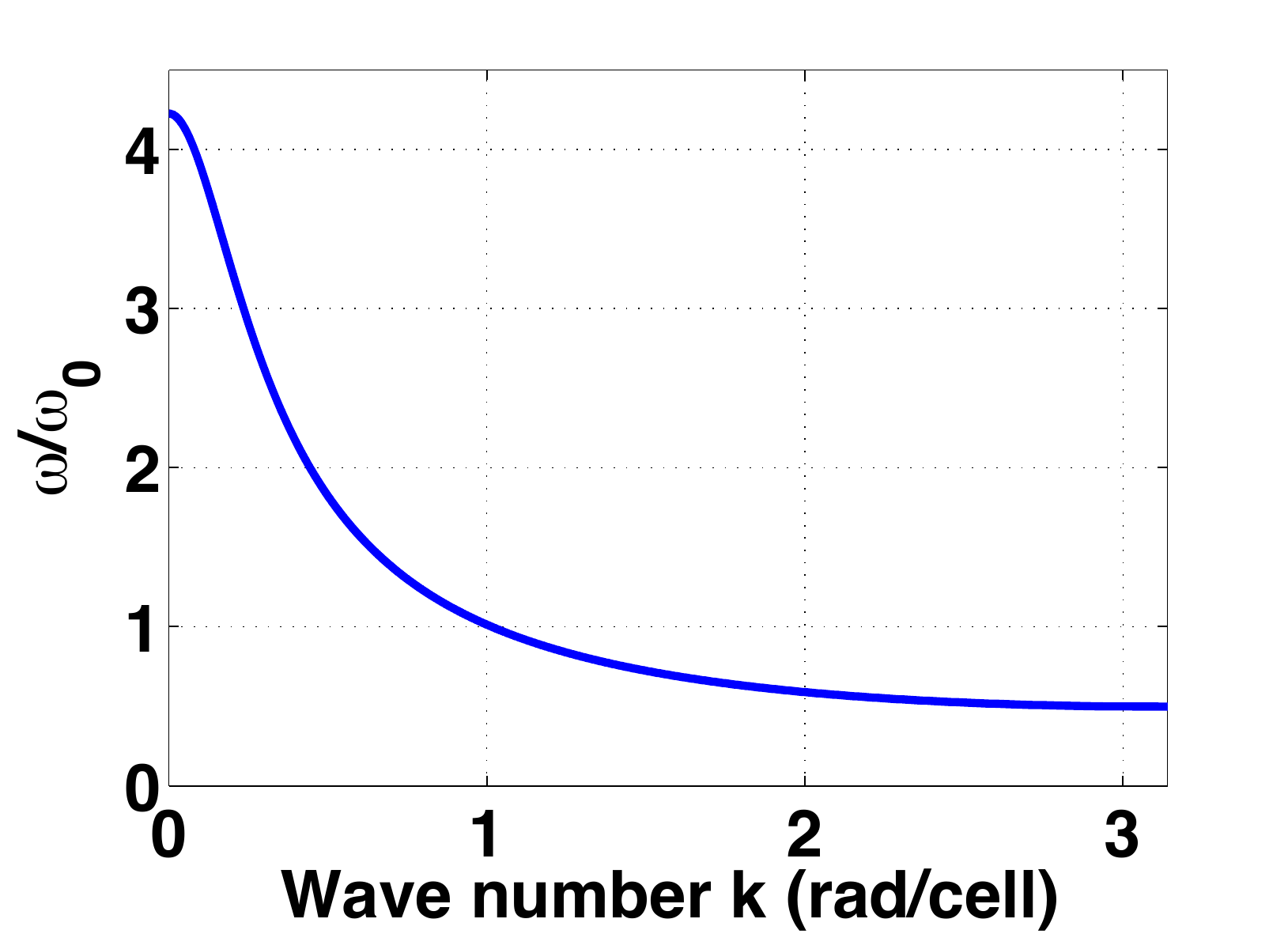}
\includegraphics[scale=0.4]{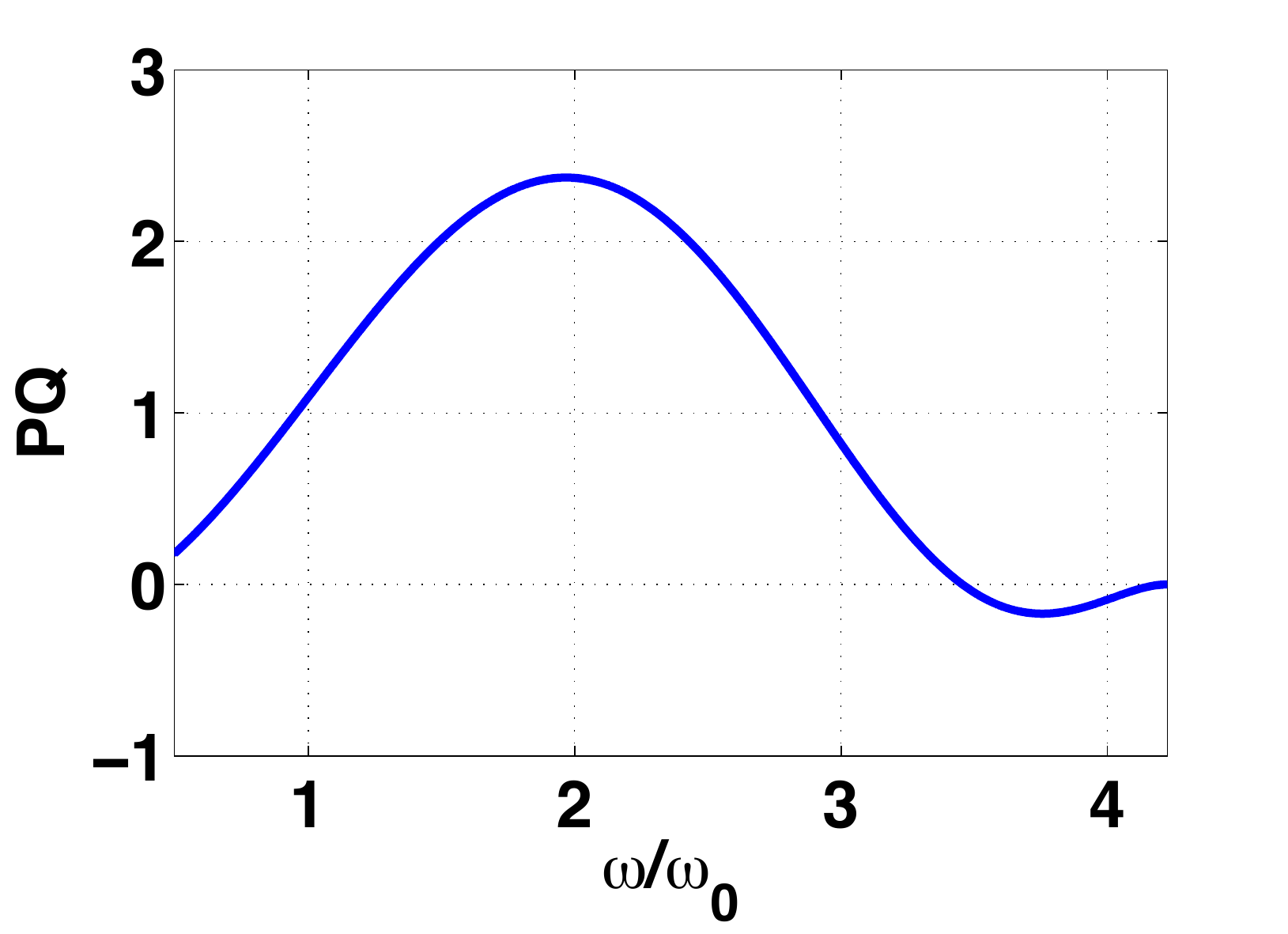}
\caption{(Color online) Left panel: the linear dispersion relation -- cf. Eq.~(\ref{disp}).
Right panel: the dependence of the 
%key 
factor $P Q$ (which determines  
%(for determining 
the focusing or defocusing nature of the model) on the frequency $\omega$;
see also the text.
When $P Q >0$, the nonlinearity is 
%deemed to be 
self-focusing
while for the opposite sign, it is self-defocusing.} \label{PQ}
\end{figure}

The first one of these, at O$(\epsilon)$, represents the linear dispersion
relation of the LHM, which is depicted in the left panel of Fig.~\ref{PQ}. 
Notice that the dispersion relation (\ref{disp}) suggests that there exist two 
cutoff angular frequencies, namely an upper one, $\omega_{max} =1/\sqrt{g}\approx 4.22$ (corresponding
to $k= 0$), and a lower one, $\omega_{min}=1/\sqrt{g+4}\approx 0.5$ (corresponding to $k = \pi$), for 
$g=0.056$; 
notice that the lower cutoff frequency is due to discreteness since, evidently, 
this frequency vanishes in the continuum limit. 

At the next order, the solvability condition yields
the group velocity (the velocity of wavepackets) $v_g=d\omega/dk$, 
which is not only distinct
from the phase velocity $v_p = \omega/k$, but also 
%. In fact, per the left-handed nature of the medium the former and the latter 
carries opposite sign, as per the left-handed nature of the medium; 
this becomes clear by the form of the dispersion relation shown in the left panel 
of Fig.~\ref{PQ}, which features a negative slope.
%While, 
Note that, at the second order, the solvability condition leads
to a vanishing contribution $V_2=0$, as is commonly the case in such
multiscale expansions. 

At the third order, we obtain the NLS 
%the nonlinear Schr{\"o}dinger 
equation for $V_1$. Its dispersion and nonlinearity coefficients, $P$ and $Q$ respectively,
depend on the frequency, but the latter is slaved to the wavenumber through the dispersion 
relation. Last, but not least, the third-order reduction/decomposition
of the solution is also derived. 

We now consider some prototypical values of the relevant parameters
motivated also in part from the experiments of~\cite{ourlars}. For instance,
for $g = 0.056$, 
%$\gamma = 16/900$ 
we present $PQ$ as function of $\omega$
in the right panel of Fig.~\ref{PQ}. In the region where the relevant 
quantity is positive, per the standard general theory
of the NLS equation~\cite{sulem,ablowitz,siambook}, the dynamics is
associated with a self-focusing scenario that should bear structures like
bright solitons, but also potentially Peregrine solitons and related
waveforms. On the other hand, when $P Q <0$, then we are in a self-defocusing
regime where, e.g., dark solitons may arise. The analytical availability
of these waveforms at the NLS level, as well as the explicit form
of the transformation allows us to express these potential solutions
in the LHM dynamics. As an aside, we note that for radio-frequency,
thin film varactors as in the case of~\cite{Meyers}, the value
of $g$ may be considerably different. However, we have verified in that
case too the existence of self-focusing and self-defocusing frequency
ranges and the persistence of the solitary wave structures explored below.
Hence, we now turn to direct numerical computations
to examine the robustness of such states (and the features
thereof) in LHMs.

%find the region support bright and dark solitons. 

%Follow the note ``LH\_TL\_draft01.pdf'' 

%\begin{eqnarray}
%&&\frac{d^2}{dt^2}(V_{n-1}-2V_n+V_{n+1})-g \frac{d^2 V_n}{dt^2}-V_n
%%\nonumber \\
%%&-&\alpha \frac{d^2}{dt^2}\left\{[(V_{n-1}-V_n)^2 - (V_n-V_{n+1})^2] \right\}
%\nonumber \\
%&+& \frac{d^2}{dt^2}\left\{ [(V_{n-1}-V_n)^3 - (V_n-V_{n+1})^3] \right\}-\mu \f%rac{d^2 V_n^3}{dt^2}=0,
%\label{modp}
%\end{eqnarray}

\section{Numerical Computations}

In the present section, we will explore %the 
different waveforms that
arise at the level of the NLS model within the realm of the LHM. 
%left handed medium. To explore the 
To do so,
we numerically integrate Eq. (\ref{modp}) using a 4th-order Runge-Kutta
method and periodic boundary conditions. 

\subsection{Bright soliton}
\begin{figure}[tbp]
\centering
\includegraphics[width=8cm]{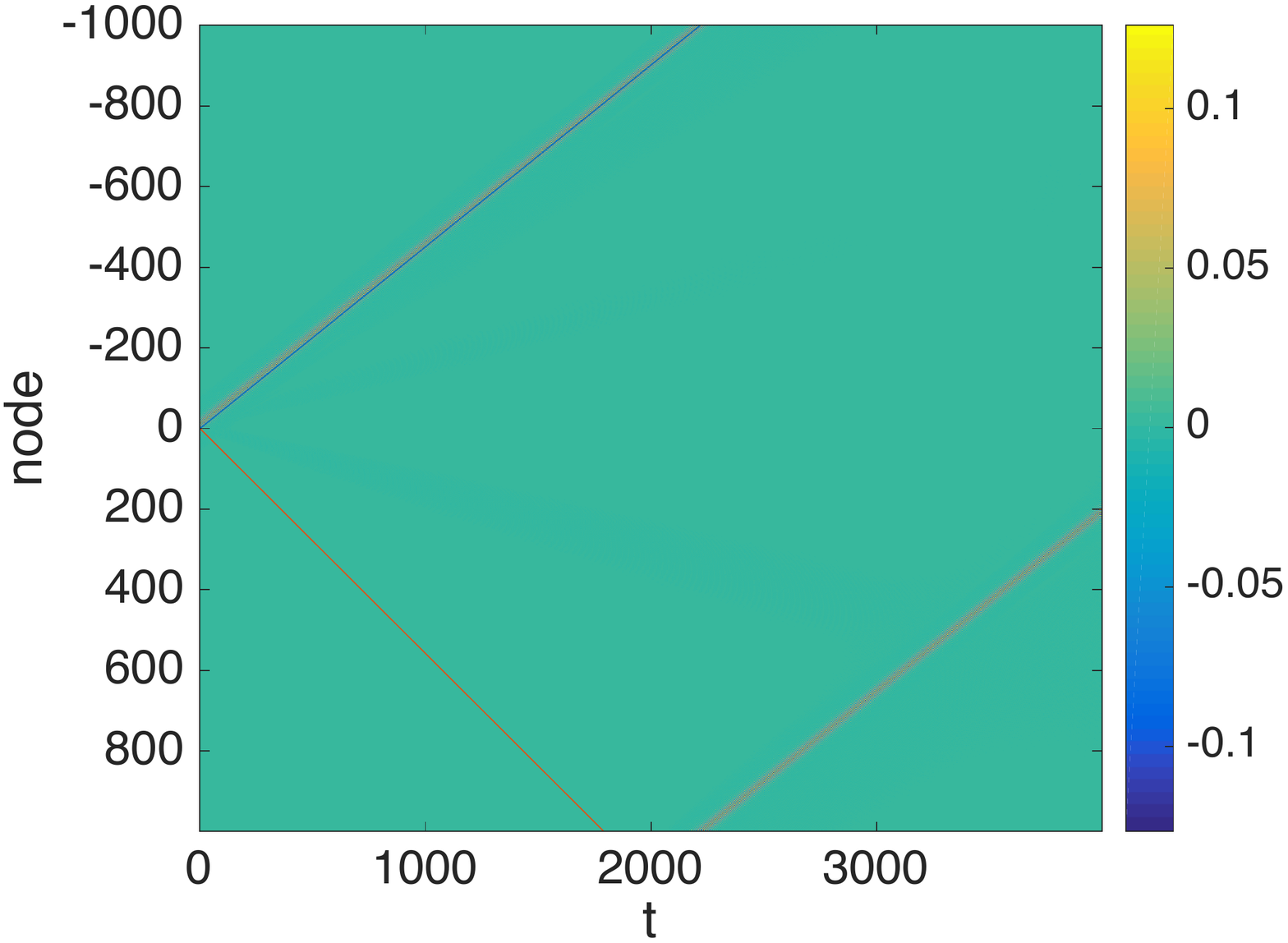}
\includegraphics[width=8cm]{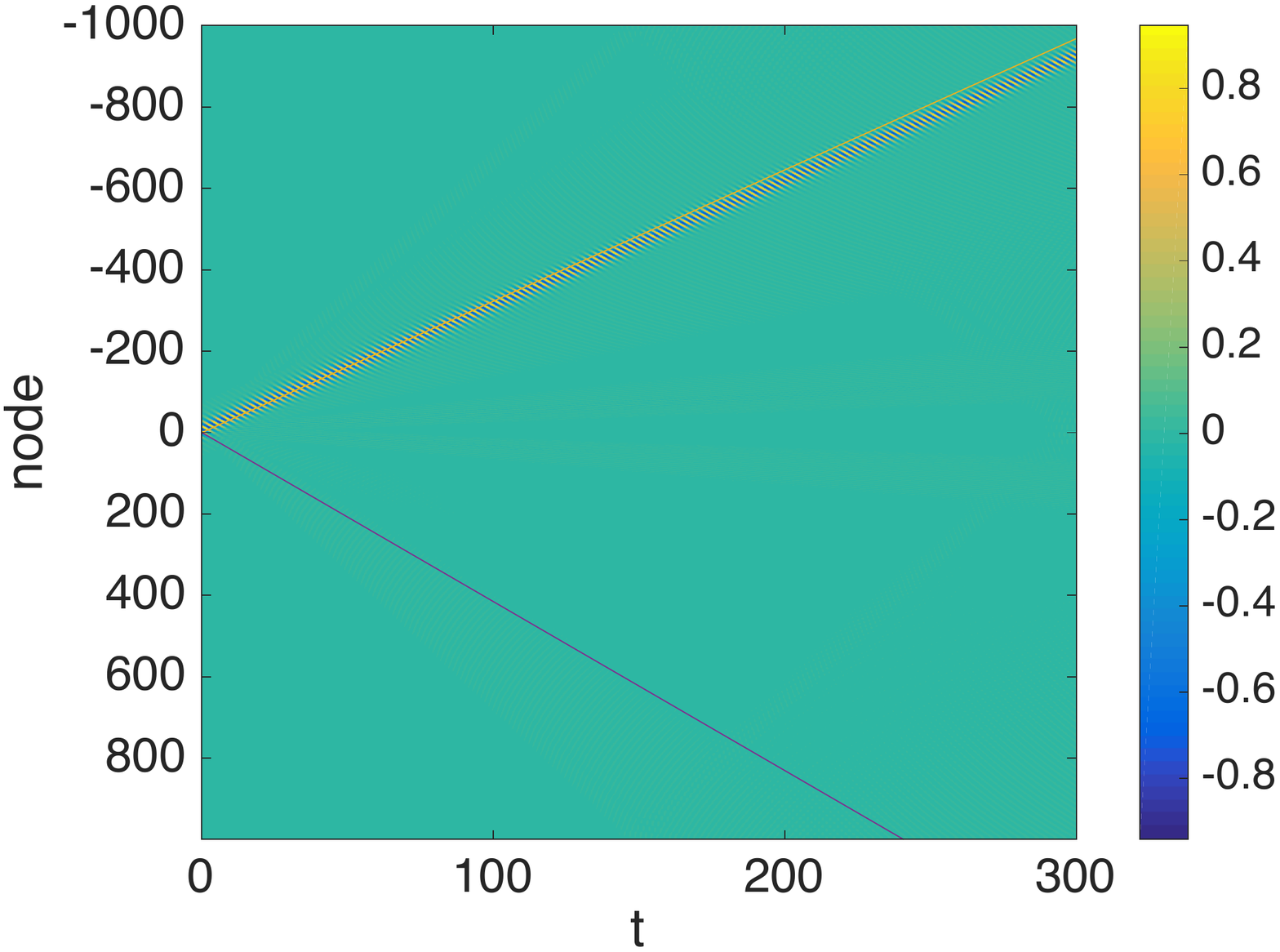}
\caption{(Color online) %In the figure, we 
Shown are contour plots depicting the space (node)-time ($t$) evolution 
%in dimensionless units, 
of a bright soliton propagating through the
left handed medium. The phase velocity $v_p=\omega/k$ is positive,
while the group 
%one 
velocity $v_g=d\omega/dk$ is negative, and are represented
by two straight lines. The soliton 
%bright solitary 
wave packet clearly propagates 
%follows the propagation 
with the prescribed group velocity.
The initial data are obtained from the  bright soliton solution 
of Eq.~(\ref{bright}) with $u_0=1$, $c=0$. For the quasi-continuum, 
%continuum, 
long-wavelength approximation, we use $\epsilon =0.1$, and 
$k =1.3823$ ($\omega \approx 0.7712$) on the left panel, $k =0.4650$ 
($\omega \approx 1.9305$) on the right panel; in both cases, $g = 0.056$.}
\label{brightEvl}
\end{figure}

It is evident from Fig.~\ref{PQ} that there exists a wide parametric
interval of frequencies, for which the effective nonlinearity of the 
NLS model is self-focusing. 
%reduction is the focusing one. 
In this case, the prototypical structure that it is relevant 
to explore is the bright soliton. 
%See Fig. (\ref{bright}).
At the level of the leading-order for the voltage, the relevant
waveform introduced on the basis of the NLS reduction has the
form:
\begin{eqnarray}
V_1 = \sqrt{\frac{2|P|}{|Q|}}u_0\sech \left(u_0(X-2c|P|T)\right)
\exp[i\left(cX+(u_0^2-c^2)|P|T\right)],
\label{bright}
\end{eqnarray}
where $u_0$ and $c$ are free O$(1)$ parameters setting the amplitude/inverse width and 
wavenumber of the soliton, respectively. 
Utilizing the above expression, and reconstructing the initial condition (of the modulated
amplitude wave, within the multiscale expansion) based on
Eqs.~(\ref{eq:ansatz})--(\ref{exp}), we can initialize the nonlinear
dynamical lattice of Eq.~(\ref{modp}) and observe the resulting
evolution presented in Fig.~\ref{brightEvl}~\footnote{Using, for instance, $L_L = 470 \mu H $, $C_0 = 800 pF$ as in~\cite{ourlars},  the time variable $t$ will be measured
in units of $\sqrt{L_L C_0} \approx 61 \mu s$ throughout our simulation
results.}. The dynamics clearly illustrates
that for different solutions of varying wavenumbers (and frequencies),
as well as amplitudes even up to order O$(1)$, we observe the extremely robust propagation
of a bright soliton through the LHM. As expected, the multiscale approximation 
is more accurate for smaller amplitudes, as observed in the left panel of
Fig.~\ref{brightEvl}. On the right panel, for larger amplitude close to
O$(1)$, 
the resulting bright soliton wavepacket tends to have smaller group
velocity than the theoretical approximation. In this case, a larger
fraction of the energy is lost to dispersive wavepacket radiation.

\begin{figure}[tbp]
\centering
\includegraphics[width=8cm, height =5cm]{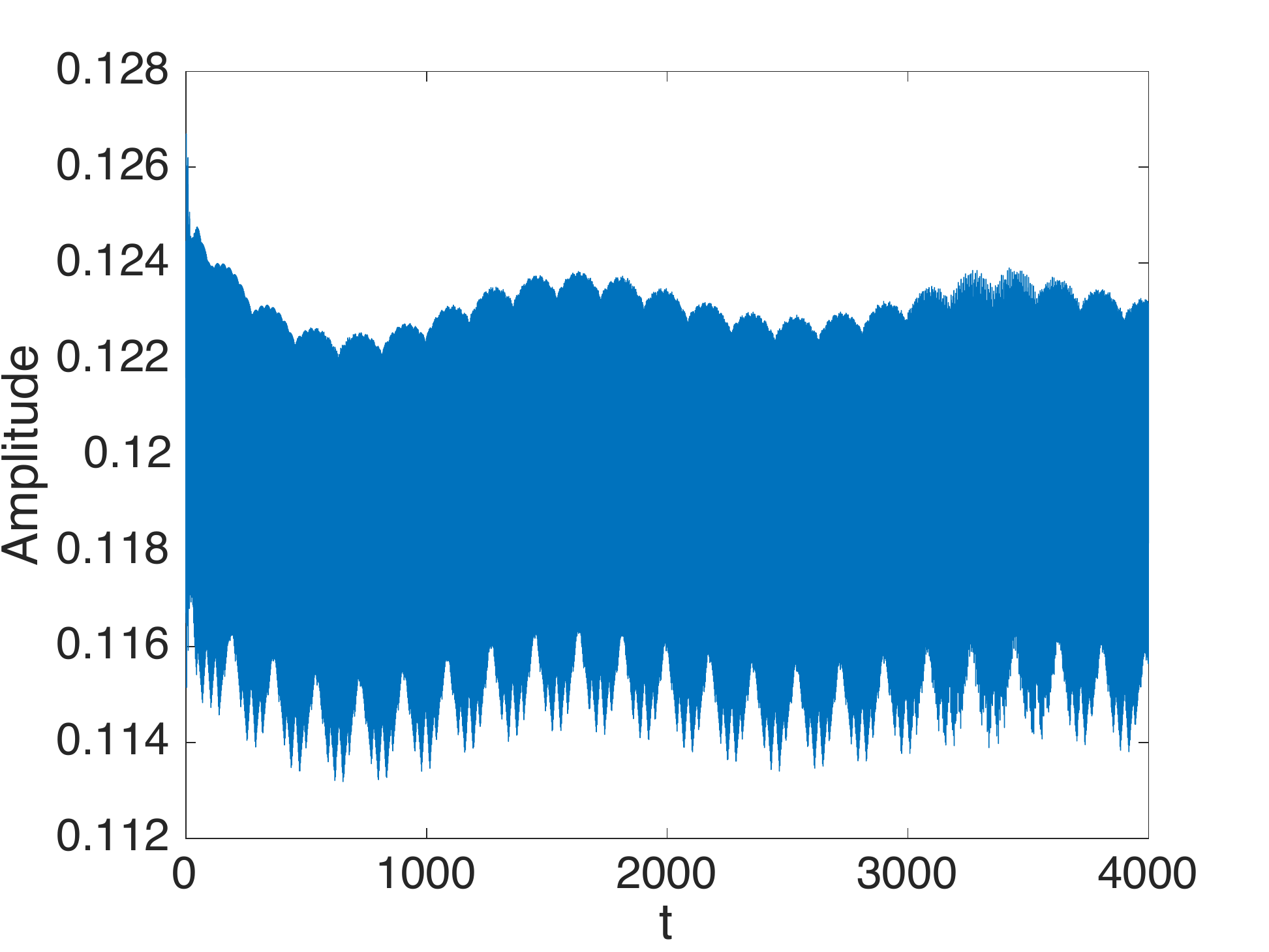}
\includegraphics[width=8cm, height = 5cm]{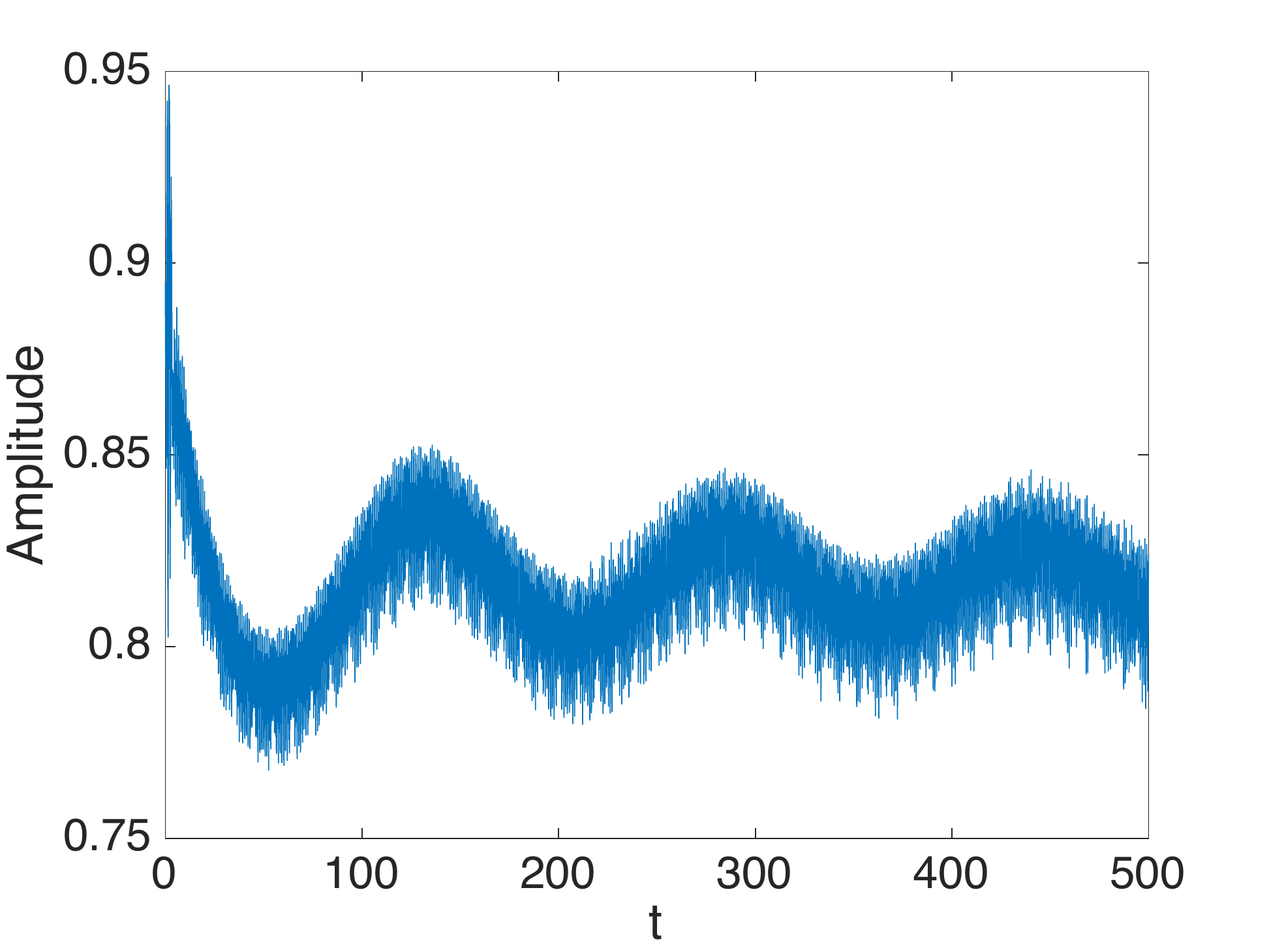}
\caption{(Color online) Amplitude (i.e., maximum voltage) of the bright soliton 
%solitary wave) 
as a function of time for Fig.~\ref{brightEvl}. Despite the
modulation induced by the LH medium, notice the robustness
of the bright soliton waveform.}
%for the left panel of Fig.~(\ref{bright2}).}
\label{brightAmp} 
\end{figure}

Figure~\ref{brightAmp} cements the relevant result by illustrating the
evolution of the amplitude of the bright soliton (i.e., the maximal
absolute value of the voltage) over time. We can see that, although the voltage
is modulated by the LH lattice, it is sufficiently robust to
be preserved under the long-time evolution. Hence, the bright soliton is an entity 
%solitary wave represents a feature
able to propagate undistorted over long distances in such transmission line metamaterials.
%lattices.

\subsection{Dark soliton}
%See Fig. (\ref{dark}).

\begin{figure}[tbp]
\centering
\includegraphics[width=12cm, height = 9cm]{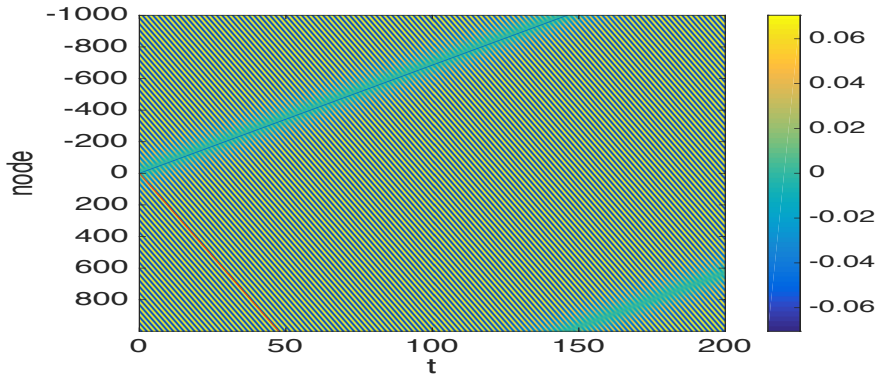}
\caption{(Color online) Dark soliton space-time contour plot evolution
for $g = 0.056$.
The initial data for the dark soliton solution are obtained
on the basis of Eqn.~(\ref{dark}) using $u_0=0.1$, $A =0$, $B=1$, $K=0$ and
$k =0.1623$ ($\omega = 3.486$), $\epsilon =0.1$ for the quasi-continuum, long-wavelength
approximation. We observe the nearly undistorted propagation of the dark soliton which is 
now supported by the defocusing NLS model.}
%structure.} 
\label{dark2}
\end{figure}

While in Fig.~\ref{PQ}, it can be observed that the interval
of frequencies considered is dominated by effective self-focusing 
dynamics, nevertheless, the quantity $P Q$ can change sign. Hence, 
it is natural to explore the potential for the formation of
dark soliton states, voltage dips on top of 
%within a finite 
a carrier wave (voltage) background. 
The relevant functional form of the dark soliton solution of Eq.~(\ref{exp}) for $PQ<0$ 
%at the leading-order of the expansion
reads:
\begin{eqnarray}
V_1 =  \sqrt{\frac{2|P|}{|Q|}}u_0 \left[B\tanh(u_0BX)+iA \right]
\exp \left[i(K X-(2u_0^2+K^2) |P|T)\right], %%dark
\label{dark}
\end{eqnarray}
where $u_0$ and $K$ is the background amplitude and wavenumber of the carrier, while 
$B$ and $A$ set the amplitude (``darkness'') and velocity of the soliton respectively 
(note that $A^2+B^2=1$). 
We have once again used Eq.~(\ref{dark}) and the long-wavelength
multiscale expansion machinery of Eqs.~(\ref{eq:ansatz})--(\ref{exp})
to construct a suitable initial condition for the dynamical lattice
of Eq.~(\ref{modp}). The result in the space-time contour plot
evolution of the voltage is shown in Fig.~\ref{dark2}.  
In this case too, although the entire background is excited,
we can observe that the voltage dip propagates essentially undistorted
over a long propagation distance, following the prescribed (through
the analysis) group velocity.

We now turn to a type of state that has not been explored in this
context, to the best of our knowledge, in a systematic,
quantitative way in any previous study, namely the Peregrine soliton.

\subsection{Peregrine soliton}
The study of solutions of the focusing NLS involving
extreme events (associated with rogue waves)
has had a long and time-honored history through
the works of Peregrine~\cite{H_Peregrine}, Kuznetsov~\cite{kuz},
Ma~\cite{ma}, Akhmediev~\cite{akh}, as well as of Dysthe and Trulsen~\cite{dt};
see also the reviews~\cite{yan_rev,solli2,onorato}.
However, it has been the recent experiments in a wide range
of areas that has significantly propelled the amount of interest
in the related wave structures. In particular, relevant experiments 
reporting observations of rogue waves 
have emerged in nonlinear optics~\cite{opt1,opt2,opt3,opt4,opt5},
mode-locked lasers~\cite{laser},
superfluid helium~\cite{He}, hydrodynamics~\cite{hydro,hydro2,hydro3},
Faraday surface ripples~\cite{fsr}, as well as parametrically driven 
capillary waves \cite{cap}, and plasmas \cite{plasma}.

In our problem, given the NLS reduction, we can utilize the Peregrine soliton solution
of the focusing NLS model in the form:
%See Fig. (\ref{Peregrine_mu0, Peregrine_mu1}).
\begin{eqnarray}
V_1 =  \sqrt{\frac{2|P|}{|Q|}}u_0 \left(1- \frac{4(1+4iu_0^2|P|T)}{1+4u_0^2X^2+16u_0^4|P|T^2} \right)\exp(i2u_0^2|P|T),
\label{Peregrine}
\end{eqnarray}
(here, as before, $u_0$ is the amplitude of the background carrier wave) 
to reconstruct the initial condition of a waveform to be 
introduced in Eq.~(\ref{modp}) via Eqs.~(\ref{eq:ansatz})--(\ref{exp}).

\begin{figure}[tbp]
\centering
\includegraphics[width=8.5cm, height =3.5cm]{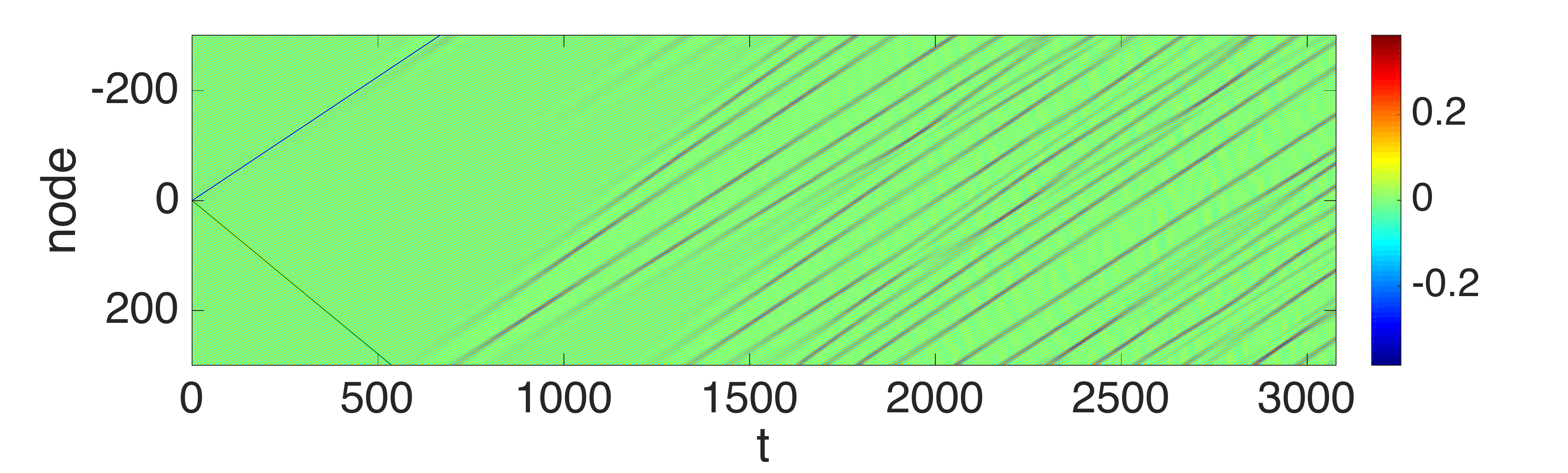}
\includegraphics[width=8.5cm, height =3.5cm]{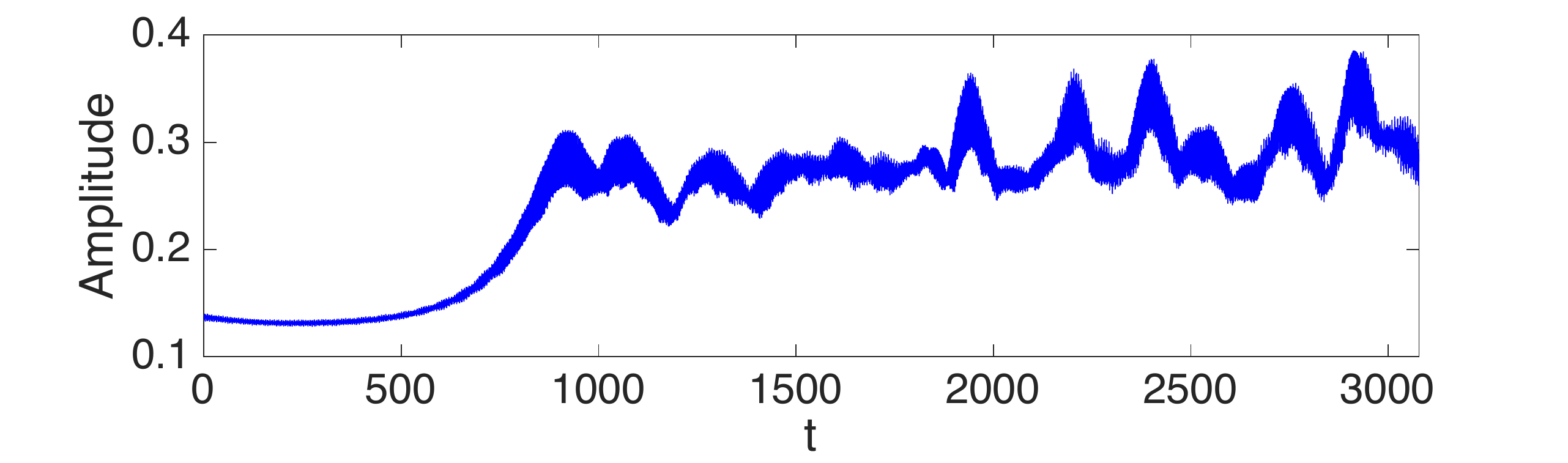}
\caption{(Color online) Evolution, through the left-handed metamaterial
lattice, of a Peregrine soliton for $\omega=0.7712$ and $g = 0.056$. For 
the quasi-continuum approximation we use $k =1.3823$ and $\epsilon =0.1$. 
The left panel shows the space-time contour plot of the lattice, while the
right panel the evolution of the maximum voltage amplitude.} \label{Pere}
\end{figure}

We initialize the relevant waveform at a time well before the formation 
of its maximum and observe its full evolution. We do this both for a smaller 
amplitude case, where the reduction should be more representative of the
true NLS dynamics, as well as for a larger amplitude one.
Figure~(\ref{Pere}) shows a Peregrine soliton example with a small amplitude; 
$\epsilon=0.1$ is used here.
%We start from a time 
%far away from the formation of the maximum of the Peregrine 
%soliton in order to observe the full evolution.
The dynamical evolution 
illustrates that the number of peaks progressively increases; i.e., while there is the
emergence of the fundamental peak associated presumably with the
Peregrine soliton, for longer times an evolution somewhat reminiscent
of modulational instability and the formation of a more complex pattern
consisting of multiple breathing solitary wave entities appears to
emerge. It is worthwhile to mention (also in connection with the
results that will follow) that the growth towards the formation
of the Peregrine soliton is not monotonic (as is expected by the exact
solution). Rather, there is a slight interval of amplitude decay
before the growth, ultimately leading to the emergence of the extreme event.
%/large amplitude formation.

%On the other hand, the amplitude go through a smaller decay, then a large growth, finally oscillating within a smaller range.

\begin{figure}[tbp]
\centering
\includegraphics[width=8cm, height = 3cm]{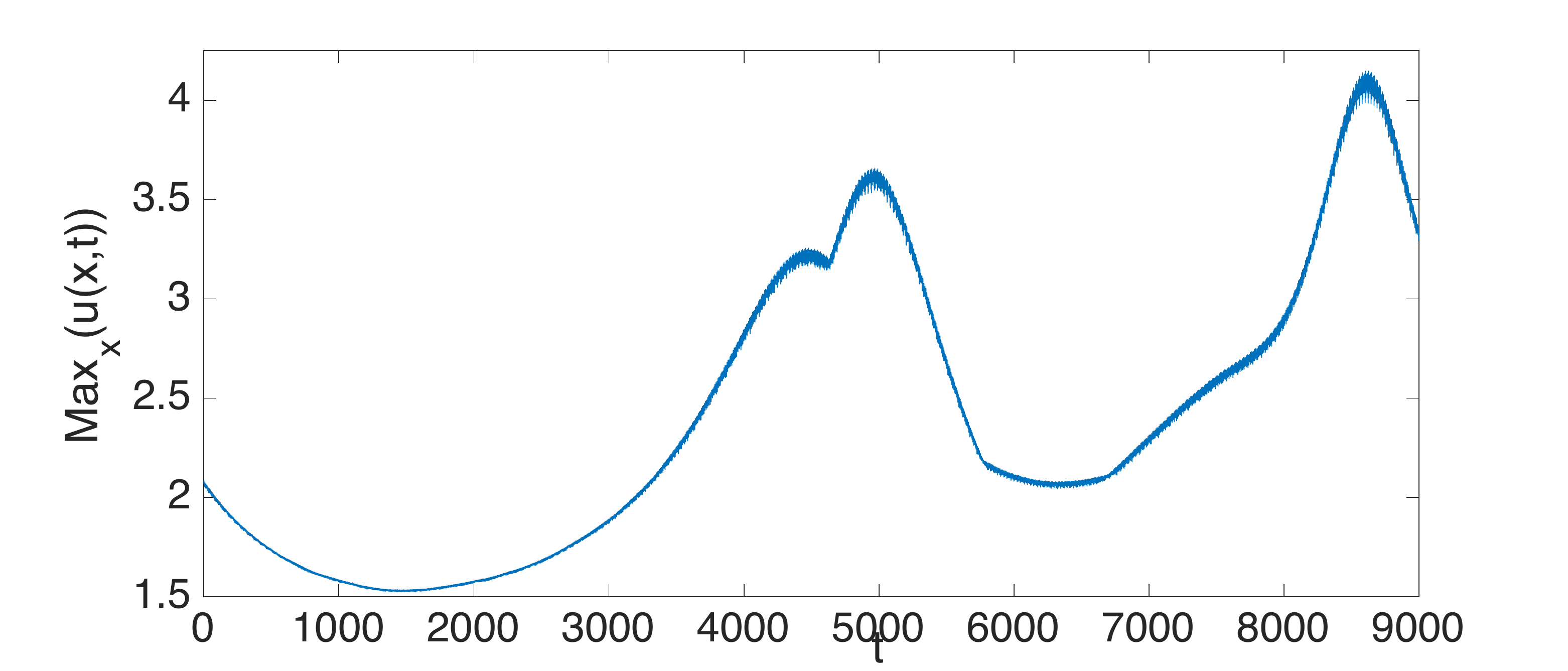}
\includegraphics[width=8cm, height = 3cm]{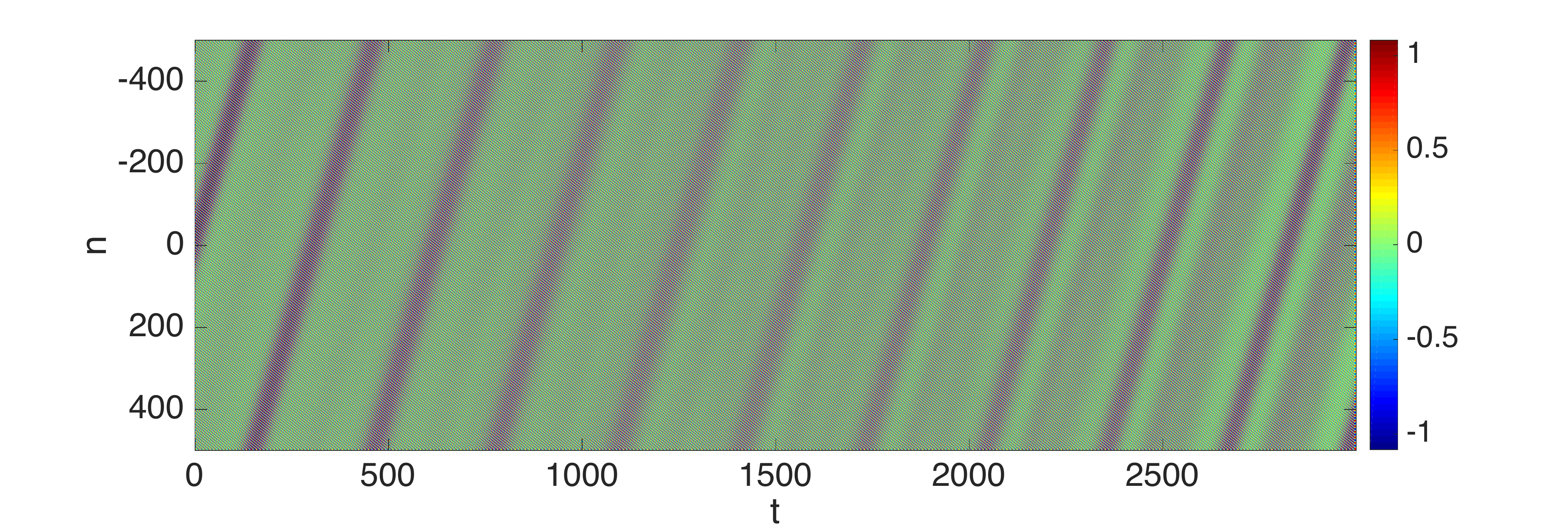}
\includegraphics[width=8cm, height = 3cm]{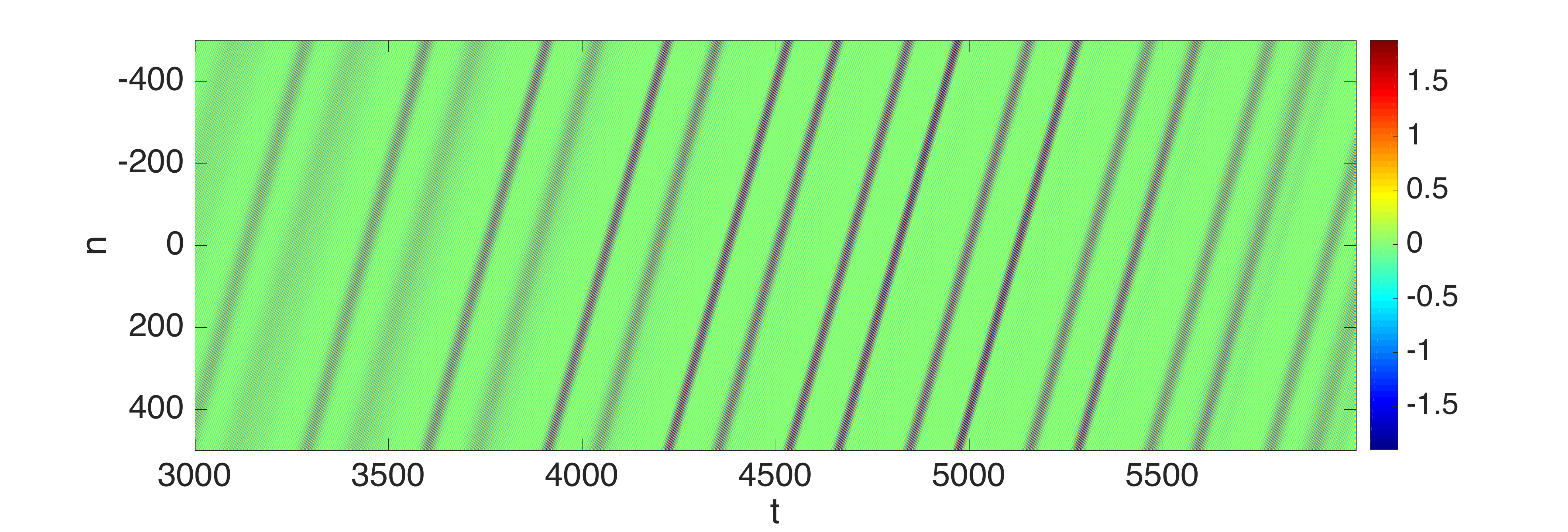}
\includegraphics[width=8cm, height = 3cm]{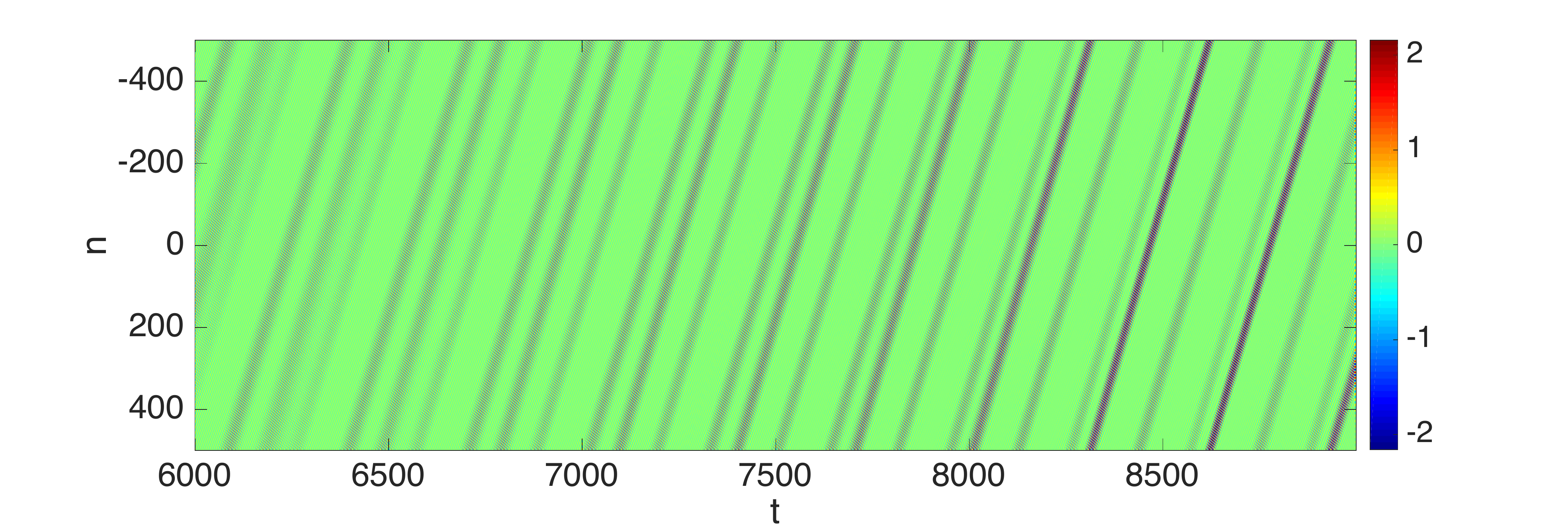}
\caption{(Color online) Evolution through the left handed metamaterial
lattice of a Peregrine soliton for  $\omega=1.926$, and $g = 0.056$.
%The leading order reduction of 
Here, Eq.~(\ref{Peregrine}) is utilized 
%through the multi-scale expansion 
to obtain the initial condition in the original variables.
%Here 
The top left panel shows the evolution of the solution's maximum.
The top right, bottom left and bottom right panels show a very long space-time
contour plot for the evolution of the voltage.
%at the level of a voltage contour plot. 
For the formation of a large-amplitude (extreme) event and the subsequent splitting
%is discussed in some 
see details in the text.}
%Initial data (left) getting from  Peregrine soliton solution of Eqn. (\ref{Peregrine}) with $u_0=1$.   For the quasi-continuum approximation we use $k =1.3823$, $\epsilon =0.05$. At $t=1000$ we get $V_n$ in the middle panel. The maximum amplitude as function of $t$ is shown on the right panel, which shows a decay then an increase when the dynamic develop two bumps.}
\label{Peregrine_mu0}
\end{figure}

Figure~\ref{Peregrine_mu0} 
%is for 
corresponds to a case of substantially larger initial
voltage, where we expect the small amplitude reduction to no longer
be valid. This case also illustrates a number of similarities and differences 
with respect to the original 
%what we expect from the standard 
NLS model. In the NLS, a monotonic 
growth of the ``bulge'' develops leading to the peak of the Peregrine
soliton (which thus seems to ``appear out of nowhere and disappear without
a trace''~\cite{akmpla}). Here, in our lattice, the growth still occurs, yet it involves a decay
stage before the growth stage leading to the peak. After the formation of the 
peak, a somewhat unconventional sequence arises in the time evolution 
of the maximum. While, that is, we expect decay anew, this decay 
occurs only briefly, with another growth stage and a sharp (in fact,
even sharper than the previous one) peak emerging. The top right panel of Fig.~\ref{Peregrine_mu0}
illustrating the space time evolution until $t=3000$ sheds light
on this feature. In particular, what happens is that the original
``wider'' waveform splits into two narrower peaks, which evolve rather
independently. At the level of the amplitude, further evolution
leads to decay and then once again to growth (the latter time developing
even higher voltage amplitudes). Once again, the contour plots of the
bottom panels for considerably larger times reveal the explanation:
in a similar way as the single wave eventually grows and splits into two, the two
subsequently proceed to split forming an additional one. This
way, the number of wave structures appears to be increasing
over time. While this is not consonant with the exact solution
of the Peregrine soliton in NLS, we should note that it is reminiscent
of an evolution leading to a progressive increase in the number
of peaks in the recent work~\cite{seas}. Furthermore, although it is
far more ordered, it carries some of the breathing characteristics
of the smaller amplitude case in Fig.~\ref{Pere}.

Thus, summarizing our findings, there exist 
%we summarize that there are 
definite similarities between the NLS reduction
and the dynamics of the LHM,
including the formation of extreme wave patterns. Nevertheless,
there are also notable differences, such as the non-monotonic growth,
or the breakup of the latter initial profile into multiple waves,
which -- especially at large amplitudes -- seems to be more complex than
what may be expected on the basis of the NLS reduction.

%\begin{figure}[tbp]
%\centering
%\includegraphics[width=5cm]{Peregrine_mu1_t0.eps}
%\includegraphics[width=5cm]{Peregrine_mu1_t1000.eps}
%\includegraphics[width=5cm]{Peregrine_mu1_Amp.eps}
%\caption{Same as in Fig. (\ref{Peregrine_mu0}), but for $\mu = 1$.} \label{Pere%grine_mu1}
%\end{figure}

\subsection{Akhmediev breathers and Kuznetsov-Ma solitons}

%However, 
As is well-known~\cite{kuz,ma,akh}, the
Peregrine soliton can be viewed as a low wavenumber or a low frequency
limit of a generalized family of solutions including on
the one hand the Akhmediev breathers and on the other Kuznetsov-Ma solitons,
respectively. Both these structures are solutions of the focusing NLS equation and 
%These NLS solutions 
can be written in a single form as:
\begin{eqnarray}
V_1 =  \sqrt{\frac{2|P|}{|Q|}} \left[1+ \frac{2(1-2a)\cosh(2 b |P|T) +i b \sinh(2 b |P|T) }{\sqrt{2a}\cos(K X) - \cosh(2b |P|T )} \right]\exp(2i|P|T),
\label{AB_PS_KMS}
\end{eqnarray}
where $b = \sqrt{8a(1-2a)}$, and  $K =2\sqrt{1-2a} $. 
For $0<a<0.5$ the solution is referred to as an Akhmediev breather, with period $2\pi/K$ in $X$. 
For $a> 0.5$, $K$ and $b$ become imaginary, thus the solution is periodic 
with period $\pi/(b|P|)$ in $T$ and Eq.~(\ref{AB_PS_KMS}) represents a Kuznetsov-Ma soliton. 
In the limit of $a\to 0.5$, these periods (spatial and temporal, respectively)
approach to $\infty$ and Eq.~(\ref{AB_PS_KMS}) has as a limiting
case the Peregrine soliton solution of NLS. Given the NLS reduction, 
we can utilize Eq.~(\ref{AB_PS_KMS}) with different values of $a$ 
to reconstruct these types of initial condition of Eq.~(\ref{modp}) 
via Eqs.~(\ref{eq:ansatz})--(\ref{exp}). 

\begin{figure}[tbp]
\centering
\includegraphics[width=8.5cm]{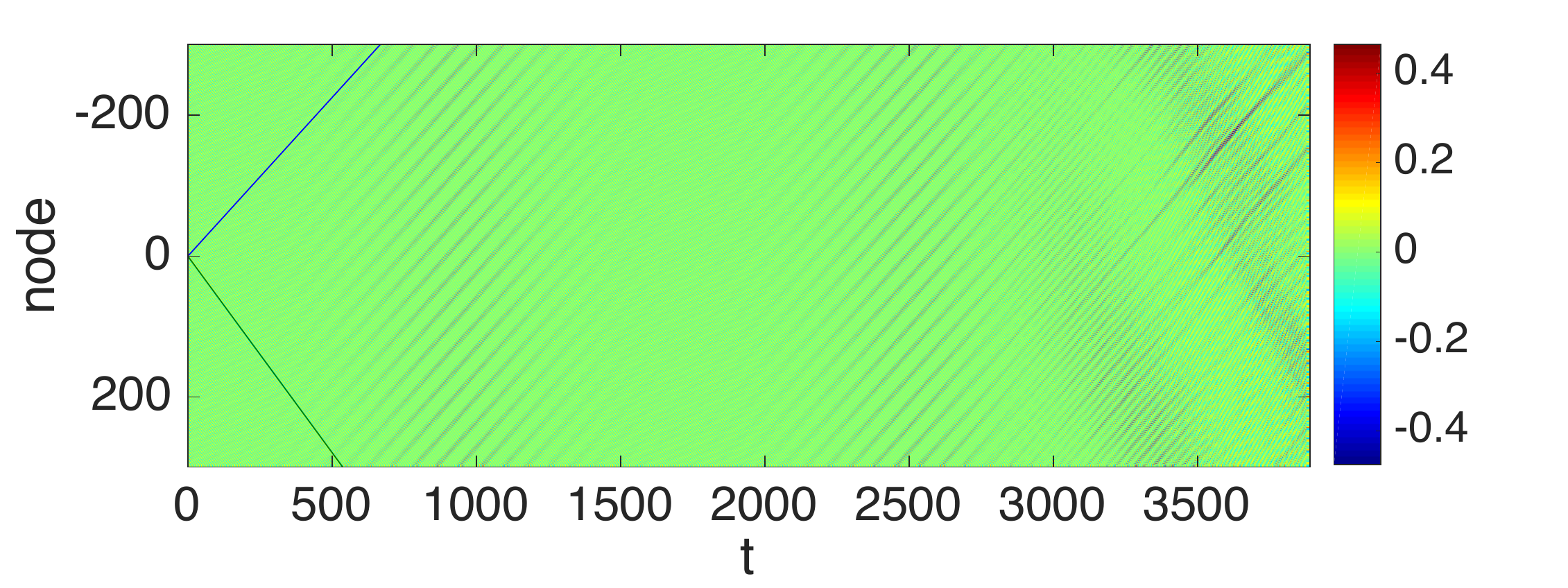}
\includegraphics[width=8.5cm]{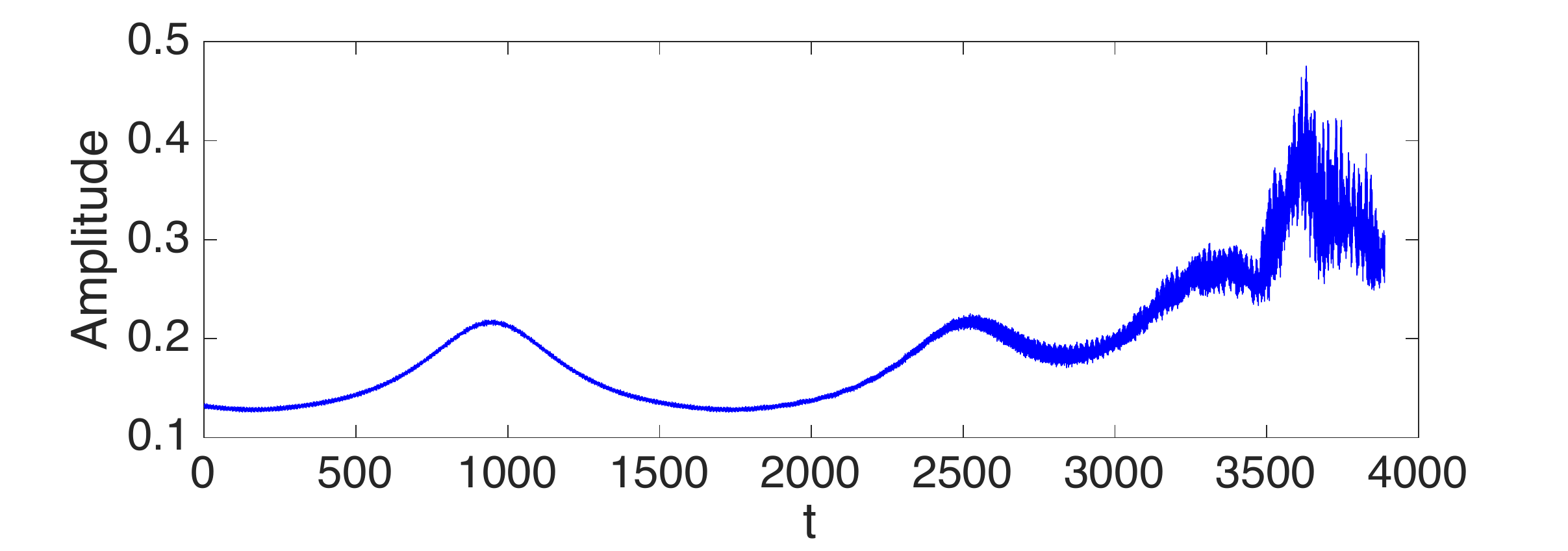}
\caption{(Color online) Evolution through the left handed metamaterial
lattice of an Akhmediev breather for $\omega=0.7712$, and $g = 0.056$.
%The leading order reduction of 
We use Eq.~(\ref{AB_PS_KMS}) with $a = 0.1131$ 
%is utilized through the multi-scale expansion 
to obtain the initial condition in the original variables. Here, 
we start from time before the formation of maximum amplitude of Akhmediev breather.
Once again, the space-time contour plot of the voltage (left) and of the maximal
evolution of the voltage amplitude over time (right) are shown.}
\label{AB_Tn}
\end{figure}

\begin{figure}[tbp]
\centering
\includegraphics[width=8.5cm]{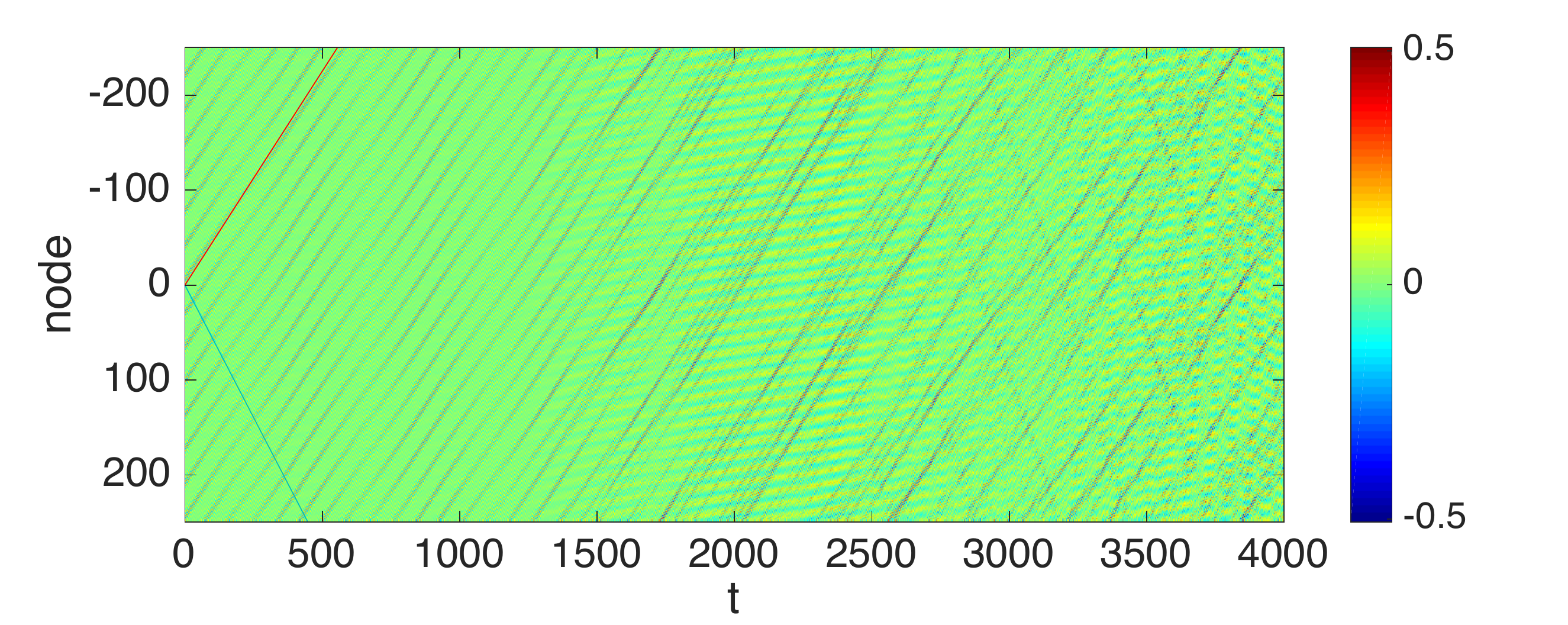}
\includegraphics[width=8.5cm]{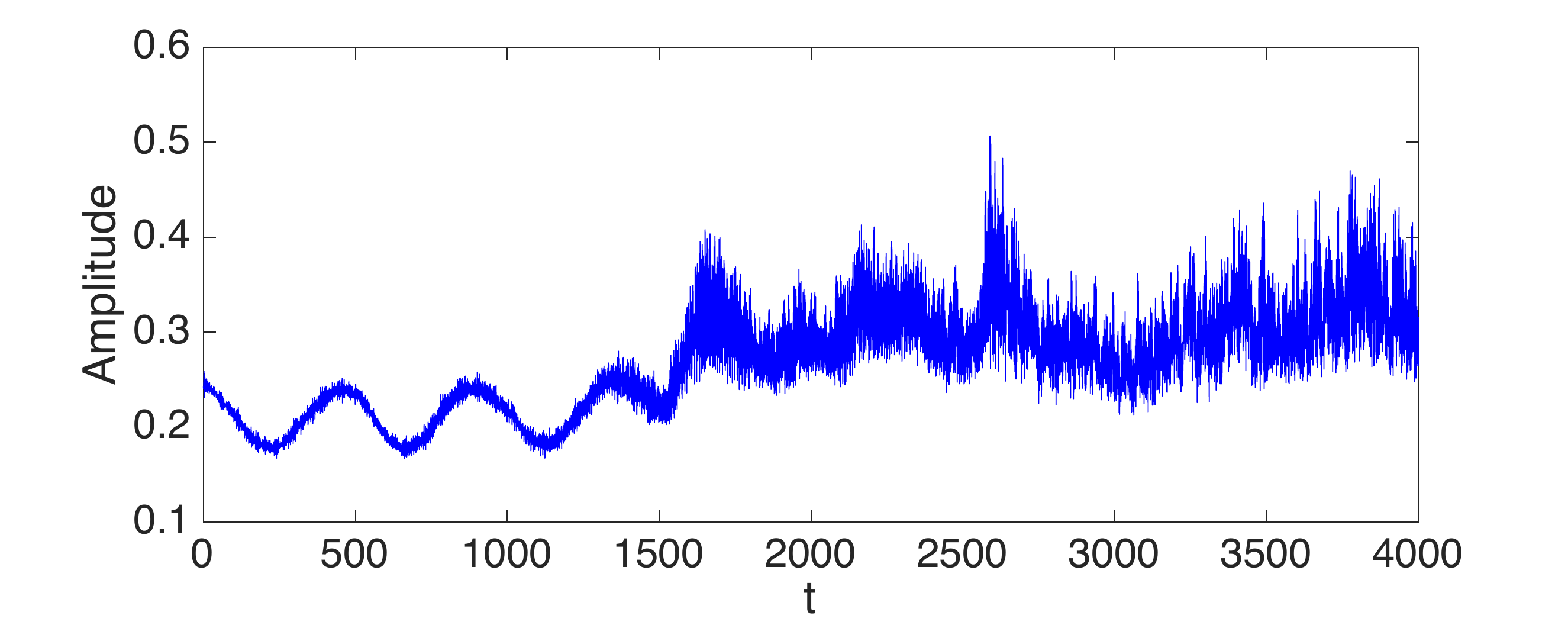}
\caption{Same as in Fig. (\ref{AB_Tn}) but for initial data using Eqn. (\ref{AB_PS_KMS}) at $T=0$.}
\label{AB}
\end{figure}

Figure~(\ref{AB_Tn}) shows the dynamics of the LHM with initial data 
taken from Eq.~(\ref{AB_PS_KMS}) for $a = 0.1131$, in the regime where 
the solution is anticipated to evolve into an Akhmediev breather. As in the 
Peregrine examples, our initialization time is before the formation of the 
maximum amplitude of the Akhmediev breather. We observe for the amplitude 
that, instead of growing to form the relevant pattern and 
subsequently decaying to a constant background forever as is prescribed by NLS, 
it oscillates until the humps become irregular.
%in 
Such a manifestation 
%that 
is, once again, somewhat reminiscent of modulational instability, 
and the subsequent formation of more highly localized waveforms. 
As a complementary simulation, we also used initial data of the LHM with $T=0$ of 
Eq.~(\ref{AB_PS_KMS}), i.e. at the maximum amplitude of the Akhmediev breather. Then 
we observe that, besides the group velocity being slightly slower than the theoretical prediction, 
the amplitude for each individual hump actually oscillates simultaneously for a while 
until around $t=1400$ with a much shorter period in Fig.~\ref{AB} than in Fig.~\ref{AB_Tn}. 
Eventually, however, in this case too the oscillatory pattern destabilizes 
and leads to an irregular profile of the energy distribution over the lattice, 
which also features occasional sharper localization phenomena.
 %than previous case. After that with the splitting of some of the humps, the pattern becomes irregular after long run.

\begin{figure}[tbp]
\centering
\includegraphics[width=9cm]{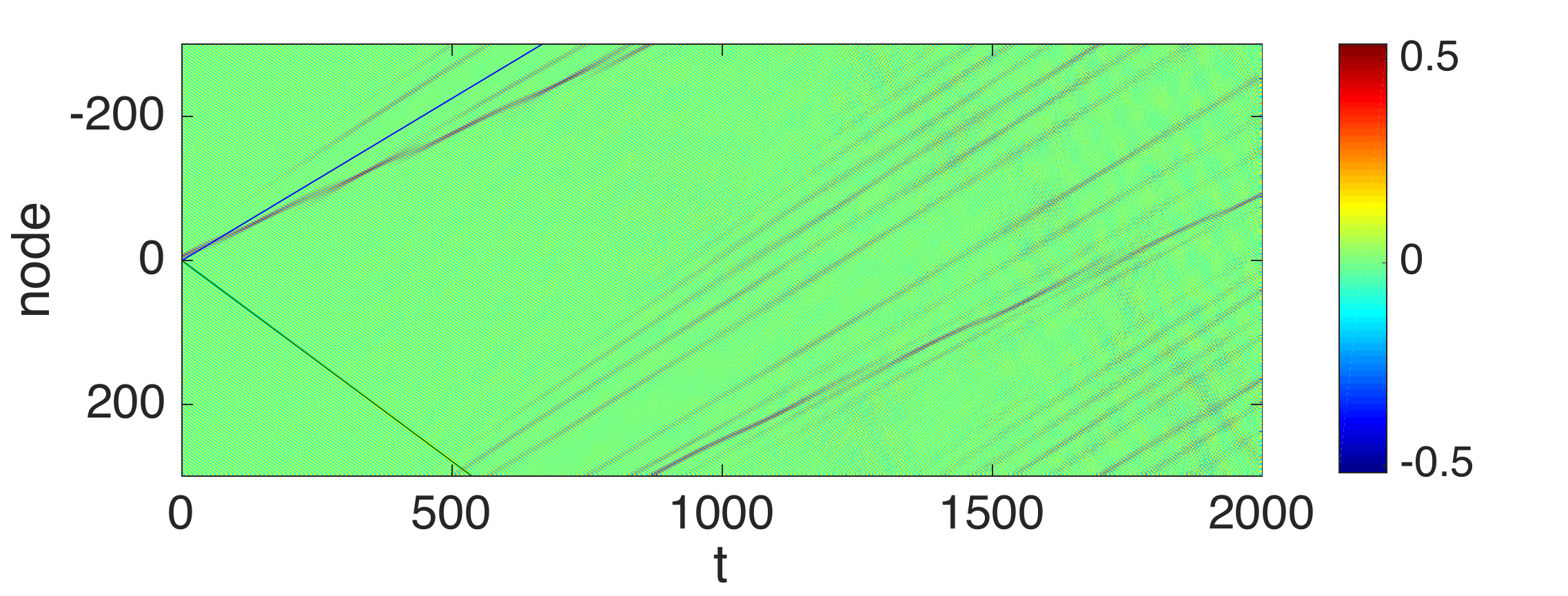}
\includegraphics[width=8cm]{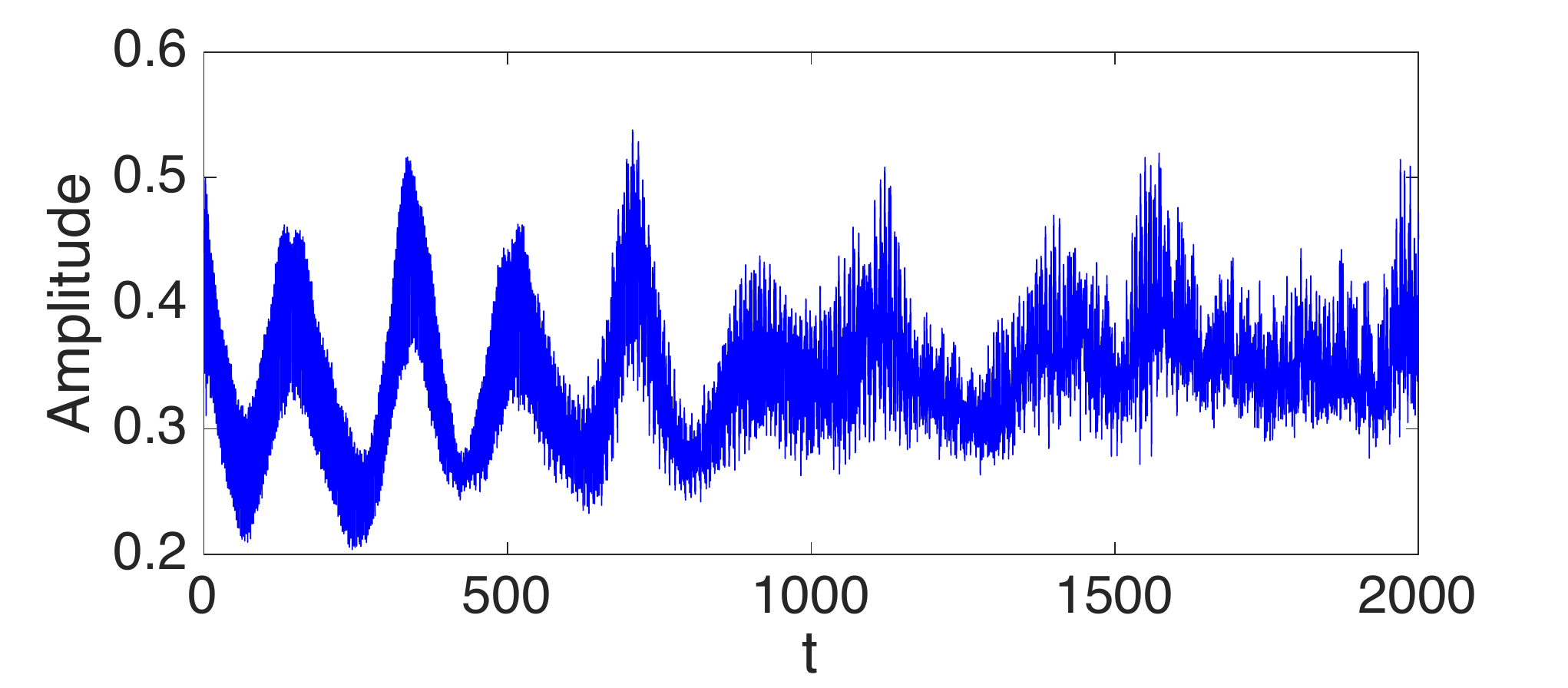}
\caption{(Color online) Evolution through the left handed metamaterial
lattice of a Kuznetsov-Ma soliton for $\omega=0.7712$ and $g = 0.056$.
%The leading order reduction of 
The solution of Eq.~(\ref{AB_PS_KMS}) with $a =0.9 $ is utilized
%through the multi-scale expansion 
to obtain the initial condition in the original variables. The same
diagnostics, namely voltage space-time contour plot (left panel) and maximal voltage vs. time 
(right panel) are depicted.
%utilized.
}
\label{KM}
\end{figure}

Figure~\ref{KM} shows the dynamical evolution of the LHM with initial data
from Eq.~(\ref{AB_PS_KMS}) at $a = 0.9$, in the regime of the Kuznetsov-Ma soliton. 
Since the Kuznetsov-Ma soliton is time periodic, this time we start from $T=0$, 
i.e. at the maximum amplitude of the Kuznetsov-Ma soliton. We observe
that the single hump splits into two humps at a very early stage, one with
group velocity considerably smaller than the theoretical prediction while the 
other one with a group velocity larger than the one suggested by the NLS
reduction. This procedure keeps cascading for the duration of our numerical 
integration, in a way once again reminiscent of the pattern formation
via modulational instability. In fact, even in the case
of the Peregrine soliton, an initialization at the maximum amplitude 
leads to the observation of similar dynamics with a splitting at an early stage. 
Besides that, as in the right panel of Fig.~\ref{KM}, we observe an oscillating amplitude, 
which is similar initially to the Kuznetsov-Ma soliton, until the
interactions with the apparently unstable background disrupt its
nearly periodic evolution.

\section{Conclusions and Future Challenges}

In the present work, we have revisited the study of left-handed transmission line  
metamaterials, motivated by the consideration of strong voltage symmetric nonlinearities 
demonstrated for epitaxially fabricated BST capacitors.
Upon introducing the relevant theoretical model, we have argued that 
its dispersive character renders it suitable for a carrier-envelope 
decomposition and an associated multiple scales reduction. This approach,
as is customary in such models, leads to a nonlinear Schr{\"o}dinger (NLS) 
equation which is a host to a diverse array of coherent waveform structures.

We illustrated that focusing, as well as defocusing nonlinearities 
can be engineered on the basis of varying the frequency (or wavenumber) of the carrier wave.
In the case of effectively self-defocusing nonlinearities, we 
observed the robust propagation of dark solitons in the system. 
In a similar way, for focusing nonlinearities, bright solitons 
%solitary waves 
were found to be generically robust. What was most interesting, 
however, was the possibility for producing extreme waveform events, 
in the form of rogue waves (Peregrine solitons, but also Akhmediev breathers
and Kuznetsov-Ma solitons) for 
such left-handed media. We observed that such events do arise 
through suitable initial conditions, motivated by the NLS 
reduction. In particular, these extreme waveforms demonstrated 
both similarities and differences
from the standard Peregrine soliton case, the differences being
the non-monotonic growth, as well as the subsequent (to the formation
of the peak) emergence of multiple peaks signaling, arguably, the
modulational instability of the background.
Similarly to the case of the Peregrine soliton, the Akhmediev breather
and the Kuznetsov-Ma soliton preserved some of their characteristics
such as the approximate spatial or temporal (respectively) periodicity,
but at the same time, they also manifested nontrivial perturbations
in both space and time, due to the modulational features of their
corresponding background.
%We discussed how this process gradually alters the single peak into
%a scenario producing three such peaks upon long time propagation. 
%{\bf Maybe, in general the constant background with hump(s) tends to produce dynamics of  more humps in a cascaded way  until eventually approach to an other background with a different level ?}

The results of the present study stimulate numerous further explorations
within this general area of the interplay of nonlinearity and left-handed media, 
especially around the subject of extreme events and rogue waves.
At the one dimensional level, it may be well worthwhile to examine
more general lattices, potentially also involving right handed
parasitic elements (as in Ref.~\cite{ourlars}), or more broadly composite
left-handed and right-handed element chains as, e.g., in Ref.~\cite{cuevas}.
A natural question is to what degree Peregrine type patterns may persist in such
settings. Another major direction for future investigations is that
of exploring the role of dimensionality. In particular, recent studies
have explored even experimentally the role of geometry (e.g., square
vs. triangular, etc.) in transmission line implementations of
two-dimensional lattices~\cite{lars2d}. It would be especially relevant
to consider left handed such media and particularly the possibility of
inducing 1D (or even more intriguingly 2D) extreme events in the latter.
Such studies are currently in progress and will be reported in future
publications.

{\bf Acknowledgments.} P.G.K., G.P.V. and D.J.F. gratefully acknowledge 
the support of QNRF Grants NPRP-8-764-1-16 and NPRP-9-329-1-067.
P.G.K. also gratefully acknowledges the
support of  NSF-PHY-1602994, the
Alexander von Humboldt Foundation, and the ERC under
FP7, Marie Curie Actions, People, International Research
Staff Exchange Scheme (IRSES-605096).
We would like to thank S. Stemmer and  R. York from the University of California at Santa Barbara for providing the BST capacitor model.

\end{document}